\documentclass[aps,prb,twocolumn,groupedaddress,showpacs,floatfix,altaffilletter,longbibliography]{revtex4-2}
\usepackage{graphicx}
\usepackage{grffile}
\usepackage{amsmath}
\usepackage{amssymb}
\usepackage{bm}
\usepackage{color}
\usepackage[dvipsnames]{xcolor}
\usepackage{amsmath,amssymb,amsfonts}
\usepackage{epsfig}
\usepackage{times}
\usepackage[colorlinks,bookmarks=false,citecolor=blue,linkcolor=red,urlcolor=blue]{hyperref}
\usepackage{gensymb}
\usepackage[toc,page]{appendix}
\usepackage{float}
\usepackage{comment}

\newcommand{\secref}[1]{Sec.~(\ref{#1})}

\newcommand{\fref}[1]{Fig.~\ref{#1}}
\newcommand{\eref}[1]{Eq.~(\ref{#1})} 
\newcommand{\sref}[1]{Sec.~(\ref{#1})}

\begin{document}

\title{Criteria and analytical results for the Pseudogap at the Van Hove point in two dimensions.}

\author{ Y.M. Vilk}
\affiliation{ 33 Weatherly Dr, Salem, MA 01970\\}
\date{\today}
\begin{abstract}
I establish the criteria and obtained analytical results for the pseudogap at the Van Hove (antinodal)  point on the Fermi surface in two dimensions. The original criterion $\xi >> \xi_{th\_db}=v_F/\pi T$ is not applicable in this case since Fermi velocity $v_F=0$. It turns out that the characteristic length for the pseudogap crossover at the Van Hove point $\xi_{th\_vh} \propto 1/T^{1/2}$,  which is significantly shorter than the one at the regular Fermi surface points $\xi_{th\_db} \propto 1/T$. In particular, $\xi_{th\_vh}$ is between one and two lattice spacing in the intermediate interaction regime of the Hubbard model. I have also identified the regime where there is still a single maximum in the spectral function, but single particle properties are abnormal. Specifically, the imaginary part of the self-energy has a minimum instead of maximum at the Fermi level and the slope of the real part of the self-energy is positive instead of negative. The important advantages of an analytical approach  is that it provides results in both the Matsubara frequency representation and in real frequencies representation. I compare Matsubara frequency results with the exact numerical results of the Monte Carlo methods in the Hubbard model and then show what they correspond to in the real frequency self-energy and spectral function.
\end{abstract}

\maketitle

\section{Introduction}
\label{sec:intro}

  Pseudogap phenomena has been observed in many quasi-two dimensional materials (for review see \cite{Kordyuk2015}). In particular it was observed in all High-Tc superconductors (for review see\cite{Vishik2018}). In the hole doped High-Tc materials it appears at temperatures above superconducting phase and gradually transitions to the full gap below phase transition. The ubiquitous feature of the pseudogap in these materials is that it appears first at points close to the Van Hove (antinodal) points on the Fermi surface. As the material cools down, the pseudogap spreads from the Van Hove point forming pseudogap arcs. Only much later, close to the phase transition, the pseudogap spreads to the diagonal of the Brillouin zone where Fermi velocity is large. Numerous theories were proposed for the pseudogap (\cite{Vilk1996, Vilk1997,varma1997non,norman1998phenomenology,franz1998phase,schmalian1999microscopic,kyung2006pseudogap,metlitski2010quantum, sordi2012pseudogap, atkinson2015charge, wu2021interplay, gauvin2022disorder, wang2023phase, dai2020modeling,ye2023crucial,sakai2023nonperturbative}. In this paper I will focus on the following ideas: the pseudogap is a precursor of the real gap in the ordered state, it is critically dependent on the dimensionality of a system (it is essentially a two-dimensional phenomenon), and it is greatly enhanced at the Van Hove points on the Fermi surface (FS). 
	
	The idea that the pseudogap should exist in two dimensions can be understood physically using the following arguments. In two dimensions (2D) the mean field phase transition to the ordered state is suppressed due to the Mermin-Wagner theorem. In the 2D anti-ferromagnetic systems the phase transition is pushed all the way down to the zero temperature. Therefore, at the phase transition the gapped state already exists with the two distinct peaks separated by the gap in the spectral function. It is thus natural to assume that this structure in the spectral function does not immediately disappear as the temperature is increased. A similar argument can be made in the superconducting case. Although in this case a finite temperature Kosterlitz-Thouless phase transition is allowed, the gap is finite at that transition \cite{moreo1992quasiparticle} and thus one can expect that the two peak structure in the spectral function remains above phase transition temperature. There is another possibility for finite temperature phase transition in quasi-two dimensional superconducting material: namely that the weak third dimension tunneling causes phase transition before Kosterlitz-Thouless phase transition occurs. This case was considered in Ref. \cite{Preosti1998} where it was shown that the pseudogap effect is very small in three dimensional systems and increases as the system becomes more quasi two-dimensional.
	
	Theoretically the pseudogap in two dimensions was predicted in the context of the repulsive 2D Hubbard model at half-filling \cite{Vilk1996}, \cite{Vilk1997}. The main result of that work was that the Fermi liquid quasi-particules are destroyed and replaced by the pseudogap when the system enters the renormalized classical (RC) regime dominated by thermal spin fluctuations. For the regular points on the FS ($v_F \neq 0 $), the pseudogap appears when correlation length $\xi$ of spin fluctuations exceeds thermal de Broglie wavelength of an electron $\xi_{th\_db}=v_F/\pi T$. Here $v_F$ is an electron velocity at the Fermi surface, $T$ is the temperature and the units are $h=1$ and $k_B=1$. I note that the positive $U>0$ Hubbard model at half-filling is equivalent to the negative $U$ Hubbard model via Lieb-Mattis canonical transformation. The antiferromagnetic ground state of the positive $U$ Hubbard model maps to superconducting ground state of $U<0$ model, the spin susceptibility with peak at antiferromagnetic Q-vector $\mathbf{Q}=(\pi,\pi)$ maps to the strongly peaked pairing susceptibility at $\mathbf{Q}=(0,0)$ and antiferromagnetic pseudogap maps to the superconducting pairing pseudogap. The criteria for the pseudogap in 2D in the superconducting context  is the same as in the antiferromagnetic case $\xi > \xi_{th\_db}=v_F/\pi T$. 
	
	Recently the pseudogap was confirmed in the 2D half-filled Hubbard model using virtually exact numerical calculations with both diagrammatic (DiagMC) \cite{vsimkovic2020extended}, \cite{Schaefer2021} and determinantal (DQMC) quantum Monte Carlo \cite{schafer2015fate}, \cite{Schaefer2021}. Although these methods have a convergence problem at low temperatures they were able to go low enough in temperature to show convincingly that the pseudogap indeed exists in the weak-to-intermediate coupling. They also show that the pseudogap first appear and is more pronounced at the Van Hove (antinodal) point on the Fermi-surface. One of the limitations of these numerical methods is that the results are available in the Matsubara frequency representation only and the analytical continuation to the real frequencies is challenging. 
		
	In this paper I establish criteria and obtained analytical results for the pseudogap at the Van Hove  point on the Fermi surface. The original criteria $\xi > \xi_{th\_db}=v_F/\pi T$ is, obviously, not applicable in this case since $v_F=0$. It turns out that the characteristic length for the pseudogap crossover at the Van Hove point $\xi_{th\_vh} \propto 1/T^{1/2}$,  which is significantly shorter than the one at the regular Fermi surface points $\xi_{th\_db} \propto 1/T$. In particular, $\xi_{th\_vh}$ is between one and two lattice spacing in the intermediate interaction regime of the Hubbard model. I have also identified the regime in which there is still a single maximum in the spectral function, but the single particle properties are anything but normal. Specifically, the imaginary part of the self-energy has a minimum instead of maximum at the Fermi level and the slope of real part of the self-energy is positive instead of negative. I call this region a false quasiparticle state. 
				
	The important advantages of an analytical method is that it allows a better understanding of the physics of the phenomena and provides results in both Matsubara frequency representation and in real frequencies representation. I compare Matsubara frequency results with the exact numerical results of the Monte Carlo methods in Hubbard model and then show what they correspond to in the real frequency self-energy and spectral function. 
	
	The paper is organized as follows. In \secref{sec:model}, I present the formalism of the approach. In \secref{sec:results} I present analytical results for the self-energy in Matsubara and real frequency representations. In \secref{sec:Monte_Carlo} I compare my Matsubara frequency results with virtually exact DiagMC method and than present real frequency results for the pseudogap regime. 
	
\section{Model}
\label{sec:model}

  In this paper I will adapt phenomenological approach. I evaluate the effect of collective mode fluctuations on the electronic self-energy of the 2D material in the one-loop approximation. The expression for the self-energy has the form:

\begin{equation} \label{eq:sEnergy_MC}
    \Sigma(\mathbf{k},ik_n) = g_{\mathbf{k}} T \sum_{iq_n} \int \frac{d^2 q}{(4\pi^2)}\chi(\mathbf{q},iq_n) \mathcal{G}_0(\mathbf{k}+\mathbf{q},ik_n+iq_n).
\end{equation}

 The above equation describes the effect of antiferromagnetic fluctuations on the self-energy $\Sigma(\mathbf{k},ik_n)$. Here $\chi(\mathbf{q},iq_n)$ is the spin susceptibility that is picked at the antiferromagnetic vector $\mathbf{q}=\mathbf{Q}=(\pi,\pi)$, $\mathcal{G}_0$ is the non-interacting Green function of an electron, and $g_{\mathbf{k}} $ is the effective coupling constant between electrons and spin fluctuations. The units are $h=1$, $k_B=1$, lattice spacing $a=1$ and in the section where I discus the Hubbard model, I set the nearest neighbor hopping parameter $t=1$.  

 The effect of the superconducting pairing fluctuation on the self-energy can be written in a similar fashion with the $\chi(\mathbf{q},iq_n)$ representing pairing susceptibility with the peak at $\mathbf{q}=(0,0)$ and the $\mathbf{k}$ and $ik_n$ in the Green function replaced by $-\mathbf{k} $ and $-ik_n$.

 Note that the Green function in the expression for the self-energy is bare rather than dressed as in the FLEX approximation. It was argued in the reference \cite{Vilk1997} that using the dressed Green function with  frequency independent vertex does not make the theory work better. In fact, it makes agreement with the Monte Carlo results worse and it does not predict the appearance of the pseudogap in two dimensions. 

 The coupling constant between electrons and collective modes $g_{\mathbf{k}}$ can, in general, depend on $\mathbf{k}$. This can be important to describe the precursor of d-wave superconductivity. In what follows the $\mathbf{k}$-dependents of coupling is not essential and I will drop the $\mathbf{k}$ index going forward. The coupling constant $g$ is actually proportional to one bare interaction $U$ and one renormalized interaction $\bar{U}$ \cite{Vilk1996}, \cite{Vilk1997}. This point will be important in \secref{sec:Monte_Carlo}, where I will compare my results with Monte Carlo benchmarks, but in the rest of the paper I will just use parameter $g$ for simplicity.

 It was argued in \cite{Vilk1997} that in the RC regime the zero Matsubara frequency term $q_n=0$ in expression \eref{eq:sEnergy_MC} is sufficient (static approximation) to describe the dominant contribution to the self-energy at the regular points on the Fermi surface (FS) $v_F \sim 1 $. However, as we will see shortly, the static approximation is not sufficient at small frequencies in the case of Van Hove point $v_F=0 $. In the latter case, one needs a more general expression that sums up all Matsubara frequencies. Using Kramers-Kronig relation for the susceptibility, summing up all Matsubara frequencies, and using standard procedures for analytical continuation $ik_n \rightarrow \omega +i0^{+}$ one can obtain  the following expression for the self-energy in real frequency representation:

\begin{equation} \label{eq:selfEnergy_RF}
  \Sigma(\mathbf{k},\omega) =g \int \frac{d^2 q d \omega'}{(4\pi^3)} \frac{\chi''(\mathbf{q},\omega') [n_B(\omega') +f(\tilde{\epsilon}(\mathbf{k}+\mathbf{q}))] }{\omega +\omega' -\tilde{\epsilon}(\mathbf{k}+\mathbf{q}) + i0^{+} }
\end{equation}

Here $\chi^{''}(\mathbf{q},\omega')$ is the imaginary part of the susceptibility,  $\tilde{\epsilon}(\mathbf{k}+\mathbf{q})$ is the energy dispersion relative to the chemical potential, $n_b$ and $f$ are boson and fermion distribution functions,respectively. The expression \eref{eq:selfEnergy_RF} is for the magnetic case. To get the the expression for pairing fluctuation case, one needs to replace $\tilde{\epsilon}(\mathbf{k}+\mathbf{q})$ with $-\tilde{\epsilon}(\mathbf{-k}+\mathbf{q})$ and $\omega'$ in the denominator with $-\omega'$.
 
 In the critical regime the correlation length growth rapidly, the susceptibility is strongly peaked at the vector $\mathbf{Q}$ ($\mathbf{Q}=0$ in the pairing case) and it can  be approximated by its asymptotic expression (Ornstein-Zernike susceptibility):

\begin{equation} \label{eq:OZ_s}
\chi_{oz}(\mathbf{q},\omega)=\frac{A}{\xi^{-2} +(\mathbf{q}-\mathbf{Q})^2-i\omega/\gamma}
\end{equation}

  Here $A$ is the prefactor or the amplitude of susceptibility, $\xi$ is the correlation length and $\gamma$ is the diffusion coefficient (Landau damping). This type of asymptotic susceptibility appears in any RPA-like  theory, whether it uses bare or renormalized interactions. However, contrary to the mean field theory, the phase transition in 2D is suppressed due to Mermin-Wagner theorem and there is a wide temperature range in which correlation length grows exponentially. In this RC regime, the characteristic energy scale of collective modes $\omega_c=\gamma \xi^{-2} $ become rapidly smaller than temperature $T$.
	
	To single out the contribution to the self-energy due to the asymptotic form of the susceptibility  \eref{eq:OZ_s}, I will add it and subtract it in the numerator of the expression \eref{eq:selfEnergy_RF}, leading to two separate contributions to the self-energy:
	
	\begin{equation} \label{eq:selfEnergy_sum}
	 \Sigma_(\mathbf{k},\omega) =\Sigma_{oz}(\mathbf{k},\omega)+\Sigma_{r}(\mathbf{k},\omega)
\end{equation}
	
	The asymptotic (Ornstein-Zernike) contribution is:

	\begin{equation} \label{eq:selfEnergy_OZ}
	 \Sigma_{oz}(\mathbf{k},\omega) =g \int \frac{d^2 q d \omega'}{(4\pi^3)} \frac{\chi''_{oz}(\mathbf{q},\omega') [n_b(\omega') +f(\tilde{\epsilon}(\mathbf{k}+\mathbf{q}))] }{\omega +\omega' -\tilde{\epsilon}(\mathbf{k}+\mathbf{q}) +i0^{+} }
\end{equation}

The regular contribution:

\begin{equation}  \label{eq:selfEnergy_R}
\begin{split}
   \Sigma_r(\mathbf{k},\omega) & =g \int \frac{d^2 q d \omega'}{(4\pi^3)} \\
	 &  \frac{[\chi''(\mathbf{q},\omega') -\chi''_{oz}(\mathbf{q},\omega')] [n_b(\omega') +f(\tilde{\epsilon}(\mathbf{k}+\mathbf{q}))] }{\omega +\omega' -\tilde{\epsilon}(\mathbf{k}+\mathbf{q}) +i0^{+} }  
	\end{split}
\end{equation}

 The $\Sigma_{oz}(\mathbf{k},\omega)$ contribution to the self-energy is primarily due to classical thermal fluctuations in the RC regime. Indeed, the frequency integral in this expression is dominated by low frequencies $\omega' \sim \omega_c=\gamma \xi^{-2} $, which is much smaller than temperature. Thus one can replace the Bose function $n_b(\omega')$ with its expansion at low frequencies  $n_b(\omega') \approx T/\omega'$ and neglect Fermi function $f(\tilde{\epsilon}(\mathbf{k}+\mathbf{q})) < 1$ in comparison with the large parameter $T/\omega_c >> 1$. The classical expression for the self-energy has the following form:

	\begin{equation} \label{eq:selfEnergy_Cl}
    \Sigma_{cl}(\mathbf{k},\omega) = T g \int \frac{d^2 q d \omega'}{(4\pi^3)} \frac{\chi''_{oz}(\mathbf{q},\omega')/\omega'} {\omega +\omega' -\tilde{\epsilon}(\mathbf{k}+\mathbf{Q}+\mathbf{q}) +i0^{+} }  
\end{equation}

 In the above expression, I used coordinate system in which Ornstein-Zernike susceptibility has maximum at $\mathbf{q}=0$. In the antiferromagnetic case it implies that the origin was shifted to $\textbf{Q}=(\pi,\pi)$. 

  The classical approximation above should be distinguished from the static classical approximation that corresponds to keeping only zero frequency term in expression \eref{eq:sEnergy_MC}. To get the latter approximation one needs to be able to neglect $\omega'$ in the denominator of \eref{eq:selfEnergy_Cl}. Then using Kramers-Kronig relation for the susceptibility, one readily arrives at the static approximation:
		
\begin{equation} \label{eq:sEnergy_st}
    \Sigma_{cl\_st}(\mathbf{k},\omega) =  g T \int \frac{d^2 q}{(2 \pi)^2}  \frac{\chi_{oz}(\mathbf{q},0)}{\omega  -\tilde{\epsilon}(\mathbf{k}+\mathbf{Q} +\mathbf{q}) +i0^{+}}
\end{equation}
	
 It was argued in \cite{Vilk1996}, \cite{Vilk1997} that for regular Fermi surface points $v_F \sim 1$ the above static approximation is sufficient in the RC regime. To see this one needs to notice that the dispersion term $\tilde{\epsilon}(\mathbf{k}+\mathbf{Q} + \mathbf{q}) $ can be expanded on small $q$ because the $q$-integrals in $\Sigma_{cl}$	expression are dominated by $q \sim \xi^{-1} << 1$. For regular Fermi surface points, the leading term of the expansion is linear $\mathbf{v_F} \mathbf{q}$ and thus of the order of $v_F \xi^{-1}$. This is much larger than $\omega' \sim \omega_c=\gamma \xi^{-2} $. For this reason $\omega'$ can be neglected in expression \eref{eq:selfEnergy_Cl} and one arrives at the static approximation for the self-energy \eref{eq:sEnergy_st}. 

  The situation is more complicated in the case of the Van Hove (antinodal) point on the Fermi surface. In this case the expansion of  $\tilde{\epsilon}(\mathbf{k}+\mathbf{Q} + \mathbf{q}) $ starts with quadratic term and thus is of order $q^2 \sim\xi^{-2}$. This is the same order as $\omega' \sim \omega_c \propto \xi^{-2} $. If electronic frequency $\omega $ is also small, then one must use a general classical expression for the self-energy \eref{eq:selfEnergy_Cl}, rather than the static approximation \eref{eq:sEnergy_st}. If one, nevertheless, would use the static approximation in this case, one would receive unphysical results: the imaginary part of $ \Sigma $ would diverge at any correlation length (see \sref{sec:Large_om} for details). The case $\omega < \gamma \xi^{-2} $  will be considered in \sref{sec:Small_om}. 
	
	When electronic frequency $\omega >>  \omega_c=\gamma \xi^{-2} $, the $\omega'$ can be neglected in the denominator of \eref{eq:selfEnergy_Cl} in comparison with $\omega$, and one again recovers the static classical approximation  \eref{eq:sEnergy_st}. This also implies that in the Matsubara frequency representation one can use the static approximation for all frequencies since even the lowest Matsubara frequency $\pi T$ is much larger than $\omega' \sim \omega_c= \gamma \xi^{-2} $.  This case will be considered in \sref{sec:Large_om}. I also note that this situation underscores challenges of numerical continuation from the Matsubara frequency results to real frequencies: a good approximation in the Matsubara representation can produce qualitatively incorrect results for small real frequencies.   
  
 Let's now turn our attention to the regular contribution to the self-energy \eref{eq:selfEnergy_R}. In the \secref{sec:Monte_Carlo}, I will compare my results with the benchmark Monte Carlo results \cite{Schaefer2021}  for the half-filled Hubbard model with the nearest neighbor hopping. To do this I need to model the regular self-energy \eref{eq:selfEnergy_R} because it gives substantial contribution at the temperatures available in the Monte Carlo calculations. I will model it by simple marginal Fermi liquid (MFL) behavior. Such behavior is expected in this model \cite{Lemay2000}, \cite{Schaefer2021}  due to perfect nesting. Importantly, it comes not from the peak in the susceptibility but from the region close to the line $(q_x,q_x)$. In particular, the MFL appears already in the second order perturbation theory which does not have strong peak in susceptibility. For this reason, I expect the MFL behavior to give the dominant contribution to the regular self-energy \eref{eq:selfEnergy_R}, from which the peak contribution was explicitly removed. I model the MFL as follows: $\Sigma''_r \propto \pi T$ for $\omega < \pi T$ ,  $\Sigma''_r \propto \omega$ for $\omega > \pi T$.

For simplicity, the results in the next section and the appendix are presented for $\mathbf{k}$ located at the FS. To get the results away from the FS, one need to replace $\omega $ with $ \omega  -\tilde{\epsilon}(\mathbf{k}+\mathbf{Q})$ in the antiferromagnetic case and $ \omega  + \tilde{\epsilon}(-\mathbf{k})$ in the pairing case.		
	
\section{Analytical results} 
\label{sec:results}

  In this section, I obtain analytical results for $\Sigma_{cl}(\mathbf{k},\omega)$ at the Van Hove point on the FS. In the antiferromagnetic case the anomalous behavior of $\Sigma_{cl}(\mathbf{k},\omega)$ occurs actually on shadow FS which obtained from the real one by the shift on the vector $\mathbf{Q}=(\pi,\pi)$. From the physical point of view, the most interesting points are "hot" spots where shadow FS and the real one intersects. I note that for the half-filled Hubbard model with nearest neighbor hopping all FS is "hot" in this sense and Van Hove (antinodal) point $\mathbf{k}=(\pi,0)$ is on the FS.
	For pairing case the maximum in susceptibility is at $\mathbf{Q}=(0,0)$ and the results here are applicable for the Van Hove point on the FS.
	
	I start by expanding on small $q \sim \xi^{-1} <<1$ the expression for electron dispersion  $\tilde{\epsilon}(\mathbf{k}_{vh}+\mathbf{Q} + \mathbf{q}) \approx w(q_x^2-q_y^2)$. Here $w=(1/2)\partial^2\tilde{\epsilon}(\mathbf{k}+\mathbf{Q})/\partial k_x^2 $. In general, the second derivatives at Van Hove point can be different in $q_x$ and $q_y$ directions by absolute value, but I assume them to be the same for simplicity. I also note that this is the case in the Hubbard model with both nearest and next to nearest neighbor hopping and Van Hove point at $\mathbf{k}_{vh}=(\pi,0)$.

	Using the above expansion for electron dispersion together with the expression \eref{eq:selfEnergy_Cl} and switching to the polar coordinates one obtains:
	
	\begin{equation} 
	\begin{split}
    \Sigma_{cl}(\mathbf{k}_{vh},\omega)& = T g \int \frac{q dq d\phi d \omega'}{(4\pi^3)} \\
		& \frac{\chi''_{oz}(\mathbf{q},\omega')/\omega'} {\omega +\omega' -w q^2 \cos(2\phi) +i0^{+} }
	 \end{split}
	\label{eq:selfEnergy_VH_g}
\end{equation}

\subsection{Results for low frequencies} 
\label{sec:Small_om}
	
Let's now turn our attention to low frequencies $ \omega << \omega_c=\gamma \xi^{-2} $. Using, the expression for asymptotic susceptibility \eref{eq:OZ_s} and integrating over $\omega'$ using delta function, one arrives at the following expression for the imaginary part of the self-energy at the Fermi level: 

\begin{equation} \label{eq:Im_selfEnergy_VH_g}
\begin{split}
    \Sigma''_{cl}(\mathbf{k}_{vh},0) & =- \frac{g A T}{\pi^2 \gamma} \int_0^\infty q dq \int_0^{\pi/2}d\phi \\
		&\frac{1}{(\xi^{-2} + q^2)^2 + w^2 q^4 \cos^2(\phi)/\gamma^2}
 \end{split}
\end{equation}

Changing variable $q^2=x \xi^{-2}$ I obtain the final result:

\begin{equation} \label{eq:Im_selfEnergy_VH_f}
    \Sigma''_{cl}(\mathbf{k}_{vh},0)  = -\frac{g A T \xi^{2}}{2\pi^2 \gamma} \sigma_{1}^{2}
\end{equation}

Where: 

\begin{equation} \label{eq:sigma_1}
 \sigma_{1}^{2} = \int_0^\infty dx \int_0^{\pi/2}d\phi \frac{1}{(x+1)^2 + w^2 x^2 \cos^2(\phi)/\gamma^2}
\end{equation}

 It is clear from expression \eref{eq:Im_selfEnergy_VH_f} that the absolute value of $\Sigma''_{cl}(\mathbf{k}_{vh},0)$ is very large in the RC regime since $\Sigma''_{cl}(\mathbf{k}_{vh},0) \propto \xi^{2} $. This has to be contrasted with the result for regular point on the FS \cite{Vilk1996}, \cite{Vilk1997} in which case $\Sigma''_{cl}(\mathbf{k}_{vh},0) $ is proportional to the first power of $\xi$ (see Appendix~\ref{app:regular_k_F}). This makes the dip in the spectral function at the Fermi level significantly more pronounced at the Van Hove point than at regular points on the FS. It also appears at higher temperatures than at regular points on the FS.

The factor $\sigma_1^2$ is of order $1$. I calculated this factor numerically for the values of $\gamma$ specific to the Hubbard model in \secref{sec:Monte_Carlo}. 

Let's now consider the real part of the self-energy. Taking real part of $\Sigma_{cl}(\mathbf{k}_{vh},\omega)$ in \eref{eq:selfEnergy_VH_g} and integrating by parts on $\omega'$ one obtains:

\begin{equation} 
	\begin{split}
    \Sigma'_{cl}(\mathbf{k}_{vh},\omega)& =  \frac{g A T}{\pi^3 \gamma } \int q dq d\phi d \omega'  \\ & \frac{2\omega'\ln|\omega +\omega' -w q^2 \cos(2\phi)|} {[(\xi^{-2} + q^2)^2 + \omega'^2/\gamma^2]^2 }
	 \end{split}
	\label{eq:selfEnergy_VH_real}
\end{equation}

The above expression can be differentiated by $\omega $ to find the slope of $\Sigma'_{cl}(\mathbf{k}_{vh},\omega)$ at $\omega =0 $. After some transformations, I obtained the following result for $\Sigma'_{cl}$ at small frequencies $\omega < \omega_c $ :

\begin{equation} 
   \Sigma'_{cl}(\mathbf{k}_{vh},\omega) = \frac{ g A T \xi^4}{\pi^3 \gamma^3 } \sigma_2^2 \omega
	\label{eq:sEnergy_real_sm}
\end{equation}

Where factor $\sigma_2^2 $ is 

\begin{equation} 
	\begin{split}
    \sigma_2^2 & =  - \int_0^{\infty} dx \int_0^{\infty} d z \int_0^{\pi/2} d\phi 
		 \frac{\ln |z^2 - w^2 x^2 \cos^{2}(\phi)|} {[(x+1)^2 + z^2/\gamma^2]^2 } \\ 
		& \left[1-\frac{4 z^2/\gamma^2}{(x+1)^2 + z^2/\gamma^2]^2}  \right]
	 \end{split}
	\label{eq:sigma2}
\end{equation}

 It is important to note that the slope of $\Sigma'_{cl}(\mathbf{k}_{vh},\omega)$ in the vicinity of the Fermi level is positive, large, and scales as $\xi^4$ . It is significantly larger than for regular point on the FS (see \secref{app:regular_k_F} ) for which the slope is proportional to $\xi^2$ \cite{Vilk1996},\cite{Vilk1997} . The slope $\partial \Sigma'_{cl}(\mathbf{k_F},\omega)/\partial \omega > 1 $ leads to two solution for the poles of the Green function (see \fref{fig:real_freq_theory_real_Sigma} ). These two poles in the Green function imply two peaks in the spectral function separated by the pseudogap (see \fref{fig:real_freq_theory_A_k_om} ). Thus the condition for precursors of quasi-particles in the ordered state are significantly more favorable at the Van Hove point than at regular point on the FS. 

 The parameter $\sigma_2^2 $ is positive and of the order of 1. To see that it is positive, note that the main contribution to the integral in \eref{eq:sigma2} come from small $ z, x <1 $ and thus the logarithm $ \ln $ in the expression is negative. I calculated the factor $\sigma_2^2 $ numerically for the values of $\gamma$ specific to the Hubbard model in \secref{sec:Monte_Carlo}. 
 
\subsection{Results for frequencies larger than $|\omega| > \omega_c$ } 
\label{sec:Large_om}

  In this section, I consider frequencies larger $|\omega| > \omega_c$. Since $\omega_c =\gamma \xi^{-2}$ is a very small parameter in the problem, this region of frequencies is very wide. As I explained in the \secref{sec:model}, one can use in this case the static approximation \eref{eq:sEnergy_st}. In particular, one can use this approximation for all fermionic Matsubara frequencies.  Substituting \eref{eq:OZ_s} in \eref{eq:sEnergy_st}, I obtain the following expression for the self-energy in the Matsubara representation:
		
\begin{equation} \label{eq:sEnergy_st_VH}
 \begin{split}
 \Sigma_{cl}(\mathbf{k}_{vh},ik_n) & =  \frac{g T A}{(2 \pi)^2} \int_0^\infty dq \int_0^{2\pi} d\phi \\   
 & \frac{1}{(\xi^{-2} +q^2) (ik_n  -w q^2 \cos(2\phi)) }
\end{split}
\end{equation}

The integrals on the angle $\phi $ and $q$ can be done exactly. The final result is:

\begin{equation} \label{eq:sEnergy_st_VH_f}
 \begin{split}
 \Sigma_{cl}(\mathbf{k}_{vh},ik_n) & =-i \frac{g T A}{4 \pi} \frac{k_n}{|k_n| \sqrt{k_n^2+w^2 \xi^{-4}}}  \\   
 &\ln \frac{|k_n|(\sqrt{k_n^2+w^2 \xi^{-4}}+|k_n|)}{w \xi^{-2}(\sqrt{k_n^2+w^2 \xi^{-4}} -w \xi^{-2}) }
\end{split}
\end{equation}

Let's now consider the case when:

\begin{equation} \label{eq:Criteria_VH}
\xi>> \xi_{th\_vh}=\left(\frac{w}{\pi T} \right)^{1/2}
\end{equation}

We will see in a moment that this is the condition for the pseudogap at the Van Hove point. Expanding over the small parameter $w \xi^{-2}/(\pi T)$ I obtain:

\begin{equation} \label{eq:sEnergy_st_VH_asymp}
 \Sigma_{cl}(\mathbf{k}_{vh},ik_n)=\frac{g  A}{2 \pi ik_n}\left[T\ln \xi +\frac{T}{2} \ln \left(\frac{2\pi T}{w}\right)\right]
\end{equation}

Taking into account that deep in the RC regime correlation length grows exponentially $\xi=\tilde{\xi}_0 \exp(T_0/T)$ the \eref{eq:sEnergy_st_VH_asymp} can be written as follows:

\begin{equation} \label{eq:sEnergy_st_VH_M_Delta}
 \Sigma_{cl}(\mathbf{k}_{vh},ik_n)=\frac{\Delta^2}{ik_n}\left[ 1+ \frac{T}{2T_0} \ln \left( \frac{2k_n}{w \tilde{\xi}_0^{-2}} \right)\right]
\end{equation}

Where gap parameter:

\begin{equation} \label{eq:Delta}
 \Delta^2=\frac{g A T_0}{2 \pi  }
\end{equation}

The $1/ik_n$ behavior of the self-energy is the signature of the pseudogap in the Matsubara representation. The temperature dependence of the prefactor $A$ is weak in the pseudogap regime \cite{Schaefer2021} and thus $\Delta^2$ depends very little on temperature as well. 

The pseudogap effect shows most dramatically in the temperature dependence of the self-energy for the the lowest Matsubara frequency $k_0=\pi T$. The self-energy in this case diverges as $T$ goes toward $0$. Indeed \eref{eq:sEnergy_st_VH_M_Delta} becomes for $k_0=\pi T$ :

\begin{equation} \label{eq:sEnergy_k_1}
 \Sigma_{cl}(\mathbf{k}_{vh},ik_0)=-i \frac{\Delta^2}{\pi T}\left[ 1+ \frac{T}{2T_0} \ln \left( \frac{2 \pi T}{w \tilde{\xi}_0^{-2}} \right)\right]
\end{equation}

The leading temperature correction to the term $\Delta^2/(\pi T) $ comes from the term proportional to $ (T/2T_0) \ln T$. This term is negative at low temperatures and twice smaller than the similar term for the regular point on the FS (see \secref{app:regular_k_F}). This explains why the self-energy at the Van Hove (antinodal) point is larger by absolute value than the one at the regular FS point (see \fref{fig:Sigm_0_T_N_AN_theory} in the \secref{sec:Monte_Carlo}). This difference, however, is pure finite temperature effect and should disappear as the temperature goes to zero.  

I now derive results for real frequencies. In order to perform analytical continuation of the expression \eref{eq:sEnergy_st_VH_f}, one needs first replace non-analytical function $|k_n|$ with equivalent analytical function $\sqrt{k_n^2}$. The analytical continuation is then obtained using standard procedure $ ik_n \rightarrow \omega + i0^{+}$. It turns out that in real frequencies representation,  there are two distinct functional forms for $\Sigma_{cl}(\mathbf{k}_{vh},\omega)$: one for $\omega < w \xi^{-2}$ and one for $\omega > w \xi^{-2}$. These two expressions connect smoothly at $\omega = w \xi^{-2}$. 

I start with presenting results for $\omega > w \xi^{-2}$. The expression for the real part of the self-energy has the form: 

\begin{equation} \label{eq:sEn_st_VH_real}
 \begin{split}
 \Sigma'_{cl}(\mathbf{k}_{vh},\omega) & =\frac{g T A}{4 \pi \sqrt{\omega^2-w^2 \xi^{-4}}}   \\   
 &\ln \frac{\omega + \sqrt{\omega^2 - w^2 \xi^{-4}}}{w \xi^{-2}}
\end{split}
\end{equation}

The expression for the imaginary part of the self-energy is:

\begin{equation} \label{eq:sEn_st_VH_imag}
 \begin{split}
 \Sigma''_{cl}(\mathbf{k}_{vh},\omega) & =-\frac{g T A}{4 \pi \sqrt{\omega^2-w^2 \xi^{-4}}}   \\   
 &\arctan \frac{\sqrt{\omega^2 - w^2 \xi^{-4}}}{w \xi^{-2}}
\end{split}
\end{equation}

To see that the above expressions describe the pseudogap let's consider frequencies $\omega >> w \xi^{-2}$. Since $w \xi^{-2} <<1 $, this covers wide region of frequencies.  In this case the expressions for $\Sigma_{cl}(\mathbf{k}_{vh},\omega)$ can be significantly simplified:

\begin{equation} \label{eq:sEn_st_VH_real_large}
 \Sigma'_{cl}(\mathbf{k}_{vh},\omega)  =\frac{g  A}{2 \pi \omega} \left( T \ln \xi + 0.5 T \ln \frac{2 \omega}{w}\right) 
\end{equation}

\begin{equation} \label{eq:sEn_st_VH_imag_larg}
 \Sigma''_{cl}(\mathbf{k}_{vh},\omega) =-\frac{g T A}{8 \omega}
\end{equation}   

The above equations can be rewritten in the RC regime using the gap parameter \eref{eq:Delta}:

\begin{equation} \label{eq:sEnergy_st_VH_RR_Delta}
 \Sigma'_{cl}(\mathbf{k}_{vh},\omega)=\frac{\Delta^2}{\omega}\left[ 1+ \frac{T}{2T_0} \ln \left( \frac{2 \omega}{w \tilde{\xi}_0^{-2}} \right)\right]
\end{equation}

\begin{equation} \label{eq:sEn_st_VH_RI_Delta}
 \Sigma''_{cl}(\mathbf{k}_{vh},\omega) =-\frac{\pi \Delta^2 T}{4 T_0 \omega}
\end{equation}   

  It is clear from the expression \eref{eq:sEnergy_st_VH_RR_Delta} that the electron's Green function has two poles (solutions of $\omega-\Sigma'(\mathbf{k}_{vh},\omega)=0$ ). These two poles in the Green function lead to two peaks in the spectral function $A(\mathbf{k}_{vh}, \omega)$: the  precursors of the quasiparticles in the ordered state. The positions of the peaks in $A(\mathbf{k}_{vh}, \omega)$ at low temperatures are given by:
	
\begin{equation} \label{eq:peaks_position}
 \omega_{peak}=\pm \Delta \sqrt{ 1+ 0.5 \frac{T}{T_0} \ln \left( \frac{2 \Delta}{w \tilde{\xi}_0^{-2}} \right)}
\end{equation}

 I note that the temperature dependent term in the above expression has a factor $0.5$,	which does not exist in the similar expression for regular point on the FS (see \secref{app:regular_k_F} ). In other words,  the shifts of the quasiparticles peaks at finite temperatures relative to their $T=0$ positions are smaller for Van Hove point than for a regular point on the FS. The width of the peaks decreases linearly with the temperature \eref{eq:sEn_st_VH_RI_Delta}.

I now consider case $\omega < w \xi^{-2}$. As was pointed out earlier,  the classical static approximation is not valid when $\omega > \gamma \xi^{-2}$. For this reason, the region $\omega < w \xi^{-2}$ is relevant only when $ \gamma < w $. The results below are for the case $ \gamma \xi^{-2} <\omega < w \xi^{-2}$.

The expression for the real part of the self-energy is:

\begin{equation} \label{eq:sEn_st_VH_real2}
 \begin{split}
 \Sigma'_{cl}(\mathbf{k}_{vh},\omega) & =\frac{g T A}{4 \pi \sqrt{w^2 \xi^{-4} -\omega^2}}   \\   
 &\arctan \frac{\sqrt{w^2 \xi^{-4 - \omega^2}}}{\omega}
\end{split}
\end{equation}

 The expression for the imaginary part of the self-energy is:

\begin{equation} \label{eq:sEn_st_VH_imag2}
 \begin{split}
 \Sigma''_{cl}(\mathbf{k}_{vh},\omega) & =-\frac{g T A}{4 \pi \sqrt{w^2 \xi^{-4} -\omega^2}}   \\   
 &\ln \frac{w \xi^{-2} + \sqrt{w^2 \xi^{-4} -\omega^2}}{\omega}
\end{split}
\end{equation}

	The above expressions lead to the unphysical results in the limit $ \omega \rightarrow 0$: the real part of the self-energy has a discontinuity at $ \omega =0$ and the imaginary part diverges as $\ln \omega $ at a finite correlation length. This underscores once again that these expressions cannot be used in the region $ \omega < \gamma \xi^{-2}$ . Instead one should use expressions derived in \secref{sec:Small_om}. 
	
	The pseudogap criteria $\xi >> \xi_{th\_vh}=(w/\pi T)^{1/2}$ was derived in this section for the Van Hove point on the FS. However, I expect it to be relevant for points on the FS  for which $\xi_{th\_vh}(T_{rc}) >\xi_{th\_db}(T_{rc})$ ($T_{rc}$ crossover temperature to the RC regime). The condition $w > v_F^2/(\pi T_{rc})$ defines the region close to the Van Hove point. It is at this region that the pseudogap should develop first. 
	
\section{Comparison with Monte Carlo results for Hubbard model } 
\label{sec:Monte_Carlo}

In this section, I compare my results with results of the virtually exact numerical DiagMC results \cite{Schaefer2021}  for the Hubbard model with the nearest neighbor hopping and at half filling. I choose this special case for comparison because all necessary data including parameters for Ornstein-Zernike susceptibility are available in this case. The energy dispersion in this model is given by:

\begin{equation} \label{eq:dispersion}
\tilde{\epsilon}(\mathbf{k})=-2(\cos(k_x) +\cos(k_y))
\end{equation}

 The Van Hove point in this model is located on the FS at the point $\mathbf{k}=(\pi,0)$. At this point the Fermi velocity is $ \partial\tilde{\epsilon}(\mathbf{k})/\partial \mathbf{k}=0$. The second derivative of the dispersion has opposite signs in $x$ and $y$ directions and its absolute value is $ |\partial^2 \tilde{\epsilon}(\mathbf{k})/\partial \mathbf{k_x}^2 |=2$. In Ref.~\citenum{Schaefer2021}, the Van Hove point was referred to as antinodal (AN) and in this section, to avoid confusion, I will refer to the Van Hove point as AN point as well. Another important point on the FS is the nodal point (N) at $\mathbf{k}=(\pi/2,\pi/2)$. At this point the Fermi velocity is maximal and the second derivative of the $\tilde{\epsilon}(\mathbf{k})$ is zero. All points on the FS can be connected by antiferromagnetic vector $\mathbf{Q}=(\pi,\pi)$ and thus all FS is affected by the antiferromagnetic fluctuations (all the FS is "hot"). 

 To compare with the benchmarks results I need parameters for the asymptotic Ornstein-Zernike (OZ) susceptibility and the coupling parameter $g$. Fortunately, all parameters necessary for evaluation of the classical contribution to the self-energy $\Sigma_{cl}$ is available in the literature. Specifically, I used the following parameters for the OZ susceptibility from Ref~\citenum{Schaefer2021}: the prefactor $A$, the correlation length $\xi$, and the diffusion coefficient (Landau damping) $\gamma$. For the coupling parameter $g$, I used the expression from the two-particle self-consistent (TPSC) approach \cite{Vilk1996},\cite{Vilk1997}. In the latter reference, the coupling was defined as $g= \tilde{g} U U_{sp}$ where the numeric factor $\tilde{g}=1/4$ and $U_{sp}$ is the renormalized interaction. The idea that one vertex has to be bare and one renormalized comes from standard diagrammatic arguments about double-counting of vertex corrections or, equivalently, from the Kadanov and Baym \cite{Kadanoff1962} formalism. The TPSC made the following assumptions: the renormalized vertex does not depend on the momentum and the frequency and its value is reduced relative to the bare interaction by the pair correlation function $g_{\uparrow \downarrow} (0)$, $ U_{sp}=g_{\uparrow \downarrow} (0) U $. In the TPSC, the pair correlation function is found self-consistently using the sum rule (Fluctuation-dissipation theorem). In the present phenomenological approach, I obtained it from the  DiagMC results \cite{Schaefer2021} for the double occupancy $g_{\uparrow \downarrow} (0) = \langle n_{\uparrow} n_{\downarrow} \rangle/ \langle n_{\uparrow} \rangle  \langle n_{\downarrow} \rangle$. The numeric factor $\tilde{g}$ is somewhat ambiguous in the TPSC because using the momentum independent vertex violates rotational invariance: the value of $\tilde{g}$ is different in the longitudinal and the transverse spin channels. In \cite{Moukouri2000} the average values of these two channels was used $ \tilde{g}=3/8$. This assumes that inaccuracies for $ \tilde{g}$ in the longitudinal and the transverse spin channels cancel one another. I found, however, in the present work that the result for the longitudinal channel works better and I used the factor $ \tilde{g}=1/4$ from the \cite{Vilk1996},\cite{Vilk1997}. 
      
 In \cite{Schaefer2021} the results for the one-loop approximation for the self-energy were used together with the numerically accurate results for the spin susceptibility in the dynamical vertex approximation (D$\Gamma$A). The results are qualitatively similar to the ones presented here but differ quantitatively. I believe the main source of difference is the difference in the coupling parameter. In \cite{Schaefer2021} the expression for coupling parameter $g$ had bare interaction $U$ for both vertices. It also used numerical factor $\tilde{g}=3/8 $.    
 
 I now turn to the discussion of the regular contribution to the self-energy \eref{eq:selfEnergy_R}. For the reasons explained in the \secref{sec:model}, I choose the marginal Fermi liquid (MFL) model. Specifically, I assume that the imaginary part of the self-energy is described by the following expressions: $\Sigma''_{r}(\mathbf{k},\omega)=b \omega$ for $\omega > \pi T$ and $\Sigma''_{r}(\mathbf{k},\omega)=b \pi T$ for $\omega <= \pi T$. This is, of course, over-simplification. For example, the second order perturbation theory predicts some frequency dependence of the $\Sigma''_{r}(\mathbf{k_N},\omega) $ for $\omega << \pi T$ \cite{Lemay2000},\cite{Schaefer2021}. 

\begin{equation}
    \text{Im }\Sigma_r(\mathbf{k}_{\text{F}}, \omega) - \text{Im }\Sigma_r(\mathbf{k}_{\text{F}}, 0) \overset{|\omega| \ll \pi T}{\sim} \begin{cases} 
      |\omega|,& \mathbf{k}_{\text{F}} \neq \mathbf{k}_{\text{N}}\\
      \sqrt{|\omega|}\,& \mathbf{k}_{\text{F}} = \mathbf{k}_{\text{N}}
   \end{cases},
\label{eqn:nesting_im}
\end{equation}
with $\mathbf{k}_{\text{N}}\!=\!(\pi/2,\pi/2)$

 The frequency variation is, however, small in comparison with the main, frequency independent, term $\Sigma''_{r}(\mathbf{k},0) $. I thus, neglected the  frequency variation of the $\Sigma''_{r} $ for $\omega \leq \pi T$. The parameter $b$ is the fitting parameter in my phenomenological model. To keep number of fitting parameters to a minimum, I choose it to be the same for all temperatures and $\mathbf{k}$ vectors. I use the value $b=-0.069$. For the calculation of the Matsubara self-energy and the real part of the self-energy, I used Kramers-Kronig relation. To do this, I needed high frequency cut off for the imaginary part of the self-energy. I choose it as $ \omega_{max}=6.4$. This parameter mostly affects high frequency behavior. The results for the regular contribution to the self-energy in the Matsubara representation, as well as, the real part of the self-energy are presented in the \secref{app:regular_se}.

 The self-energy is calculated as the sum of the classical and regular contributions \eref{eq:selfEnergy_sum}. Figures below show comparison of my model with the benchmark DiagMC results \cite{Schaefer2021} for $U=2$. Since, my analytical results for the classical contribution to the self-energy are valid only in the RC regime $\gamma \xi^{-2} < T$, I limit comparison to $T \leq T_{rc} $. For $U=2$ it was found that $T_{rc}=0.1$ \cite{Schaefer2021}. 

\subsection{Comparison with Matsubara frequency results} 
\label{sec:Comp_Mats_Freq}
 
\begin{figure}
    \centering
    \includegraphics[width=\columnwidth]{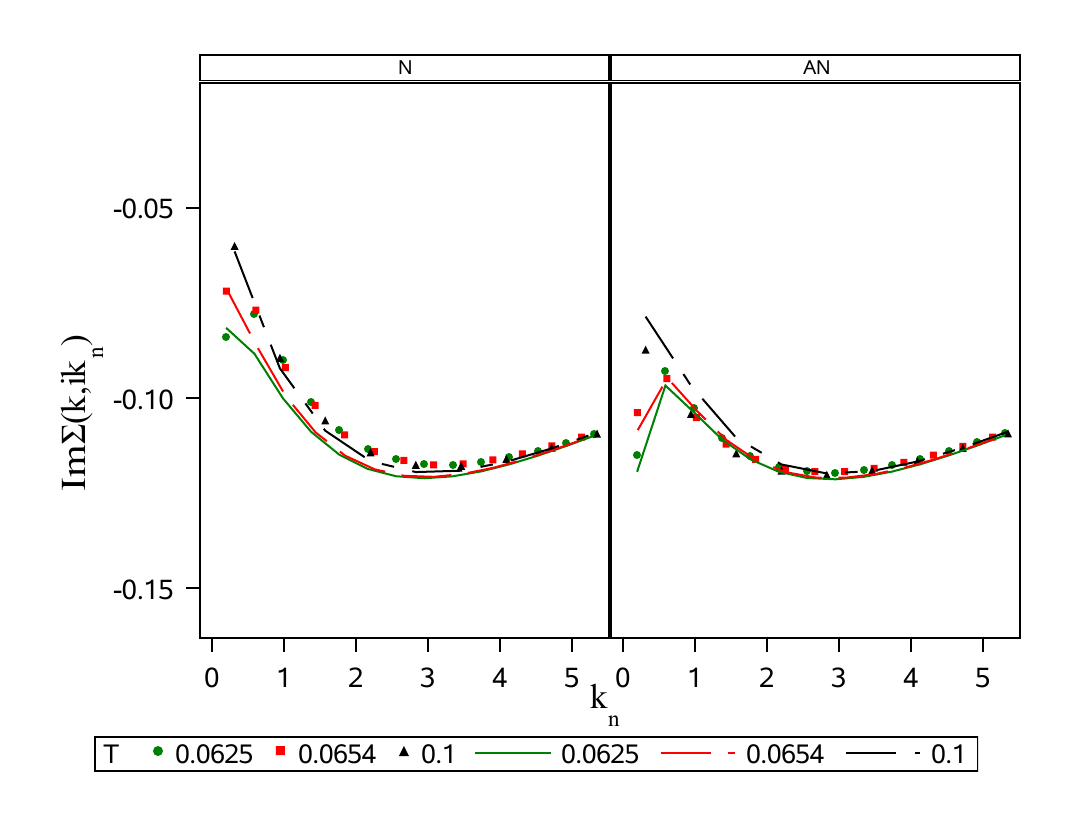} 
    \caption{Imaginary part of the self-energy for antinode (right) and node (left) as a function of Matsubara frequencies for the numerically exact DiagMC (symbols) and the present theory (lines). The drop of the self-energy at the smallest Matsubara frequency is the signature of the pseudogap.}
    \label{fig:Sigm_om_DiagMC_theory}
\end{figure}

 The \fref{fig:Sigm_om_DiagMC_theory} show the imaginary part of the Matsubara self-energy as the function of the Matsubara frequencies for three temperatures and two points on the FS: nodal and antinodal. The agreement with the present model is fairly good. The drop of the self-energy at the smallest Matsubara frequency is the signature of the pseudogap. It can be understood using the asymptotic expression \eref{eq:sEnergy_st_VH_M_Delta} for the classical contribution to the Matsubara self-energy. The latter predicts that the absolute value of the Matsubara self-energy  increases with the decrease of the Matsubara frequencies $ \Sigma_{cl} \approx \Delta^2/i k_n$. The regular contribution to the self-energy behaves in the opposite fashion: the absolute value $ \Sigma_{r} $ decreases with decreasing frequency. Thus there is a competition between two contributions. For the two lowest temperatures in the \fref{fig:Sigm_om_DiagMC_theory}, only the self-energy at the lowest Matsubara frequency clearly shows the pseudogap-like behavior. I predict that this behavior will spread to higher Matsubara frequencies when the temperature is lowered. The figure also clearly shows that the pseudogap effect comes earlier and is more pronounced for antinodal point. As I pointed out earlier, the pseudogap opens up when the correlation length is significantly larger than the characteristic thermal wave length. For the nodal point this is the de Broglie wavelength of an electron $\xi_{th\_db}=v_F/\pi T$ and for the antinodal point it is the Van Hove characteristic length $\xi_{th\_vh}=w/(\pi T)^{(1/2)}$. At low temperatures the latter is significantly shorter than the former. Specifically, for $U=2$, $T_{rc}=0.1 $ : $\xi_{th\_db} = 9$ and $\xi_{th\_vh}=1.78$. This explains, why the pseudogap appears earlier at the antinodal point than at the nodal point. 

\begin{figure}
    \centering
    \includegraphics[width=\columnwidth]{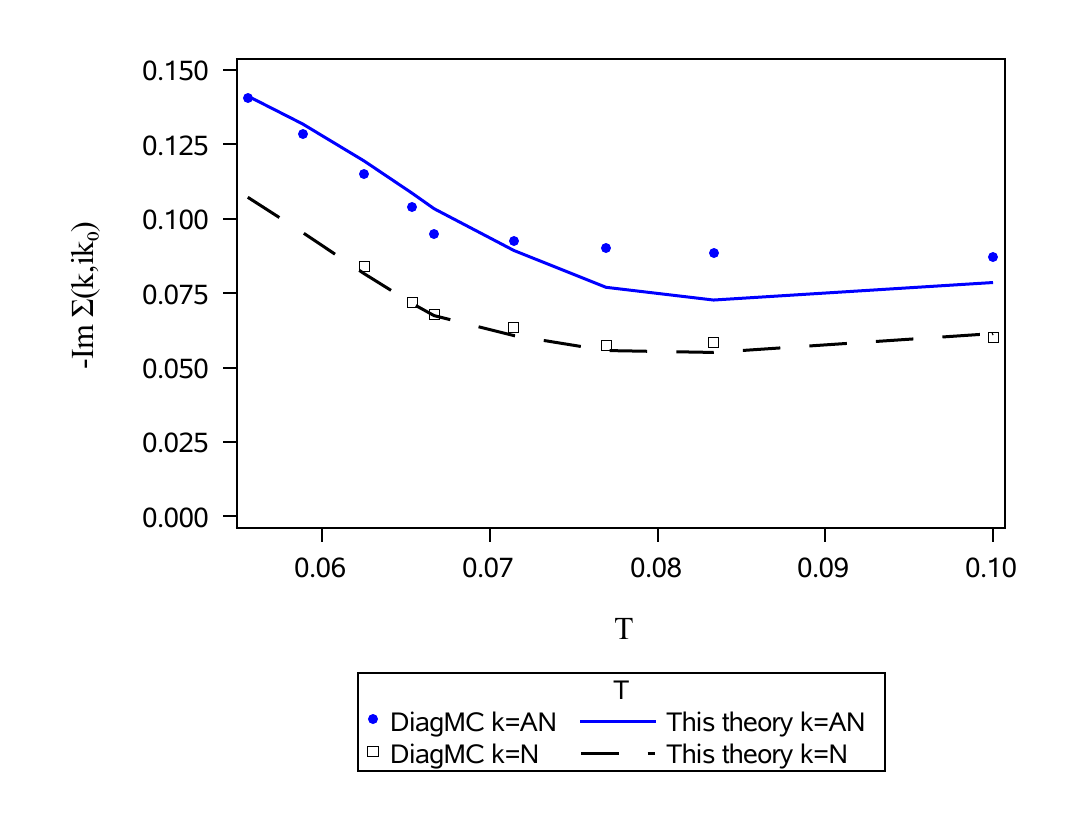} 
    \caption{Minus imaginary part of the self-energy at the first Matsubara frequency for antinode and node as a function of temperature for the numerically exact DiagMC (symbols) and the present theory (lines). The rapid increase of this quantity with decreasing temperature is a strong evidence of the pseudogap. The theory predicts divergence of the self-energy  as $T \rightarrow 0$ (see \eref{eq:sEnergy_k_1}.} 
    \label{fig:Sigm_0_T_N_AN_theory}
\end{figure}

 The \fref{fig:Sigm_0_T_N_AN_theory} shows the temperature dependence of the $-\mathrm{Im} \Sigma(\mathbf{k},i k_0)$ for the lowest Matsubara frequency $k_0 = \pi T $. The figure presents, probably, the most clear evidence of the pseudogap in the Matsubara representation because this quantity is expected to diverge as $T \rightarrow 0$ (see \eref{eq:sEnergy_k_1}. Indeed, the figure shows this tendency of the self-energy as temperature lowers. One can also notice, that the absolute value of the self-energy $-\mathrm{Im} \Sigma(\mathbf{k},i k_0)$ is larger for the antinodal point than for the nodal point. This is expected from the asymptotic behavior of $-\mathrm{Im} \Sigma(\mathbf{k},i k_0)$ for the antinodal and the nodal points, \eref{eq:sEnergy_k_1} and \eref{eq:sEnergy_st_R_M_Delta}. The leading term $ \Delta^2/(\pi T)$ is the same for both FS points. However, the temperature correction to this term is $ \propto T \ln T$, has opposite sign to the main term, and is smaller by the factor of $2$ for the antinodal point. I note that this difference between values of $-\mathrm{Im} \Sigma(\mathbf{k},i k_0)$ at different points on the FS is expected to shrink to zero as $T \rightarrow 0$.

\subsection{Real frequency results} 
\label{sec:Real_frequency}

  In this section I present real frequency results for the same parameters of the model as above. I have analytical results for $\omega << \gamma \xi^{-2}$ and $\omega >> \gamma \xi^{-2}$. I will interpolate between these two results using the following procedure: I extended results for $\omega << \gamma \xi^{-2}$ until they intersect with the results for  $\omega >> \gamma \xi^{-2}$ at some frequency $\omega_i \sim \gamma \xi^{-2} $. The interpolation procedure leads to some artifacts at frequencies $\omega \leq \omega_i$. For example, the imaginary part of the self-energy is constant for $\omega \leq \omega_i$. These artifacts at small frequencies does not affect the main results.
	
\begin{figure}
    \centering
    \includegraphics[width=\columnwidth]{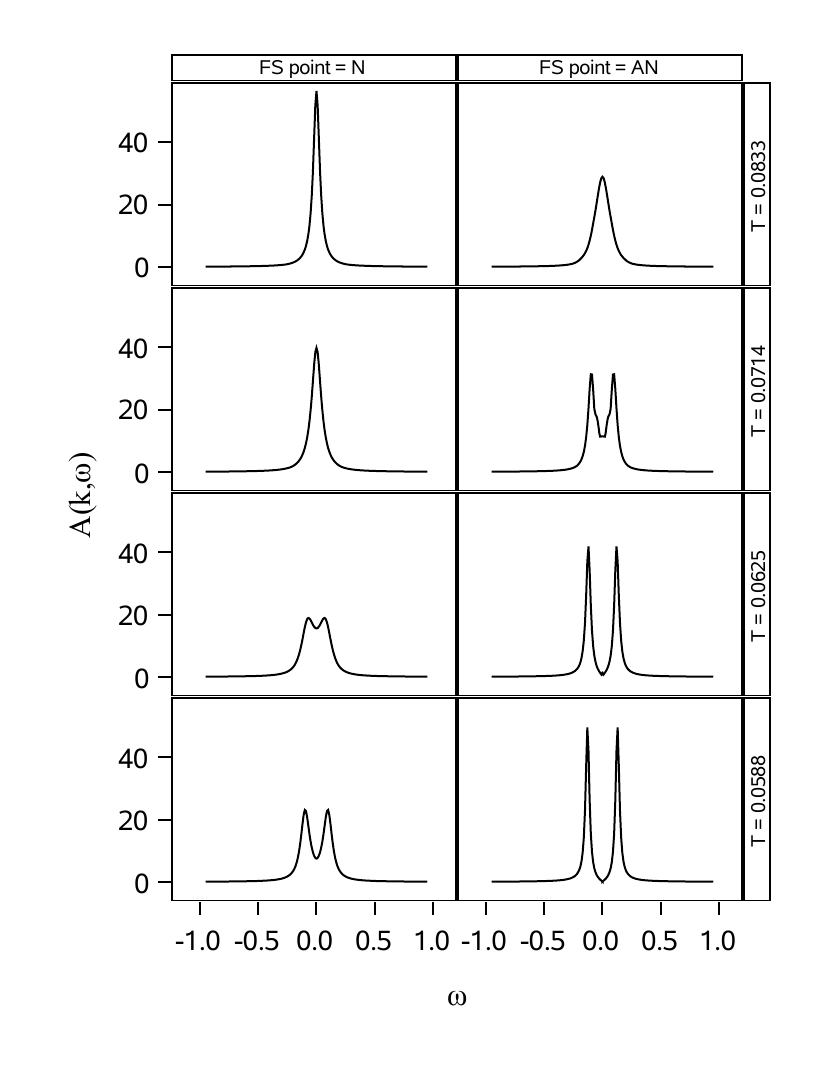} 
    \caption{Evolution of the spectral function $A(\mathbf{k}, \omega) $ with the temperature for nodal (left) and antinodal (right) points. The pseudogap appears first and is more pronounced at the antinodal point. The single maximum at $T=0.0833$ for AN point is an example of the false quasiparticle (see text for details). }
    \label{fig:real_freq_theory_A_k_om}
\end{figure}

\begin{figure}
    \centering
    \includegraphics[width=\columnwidth]{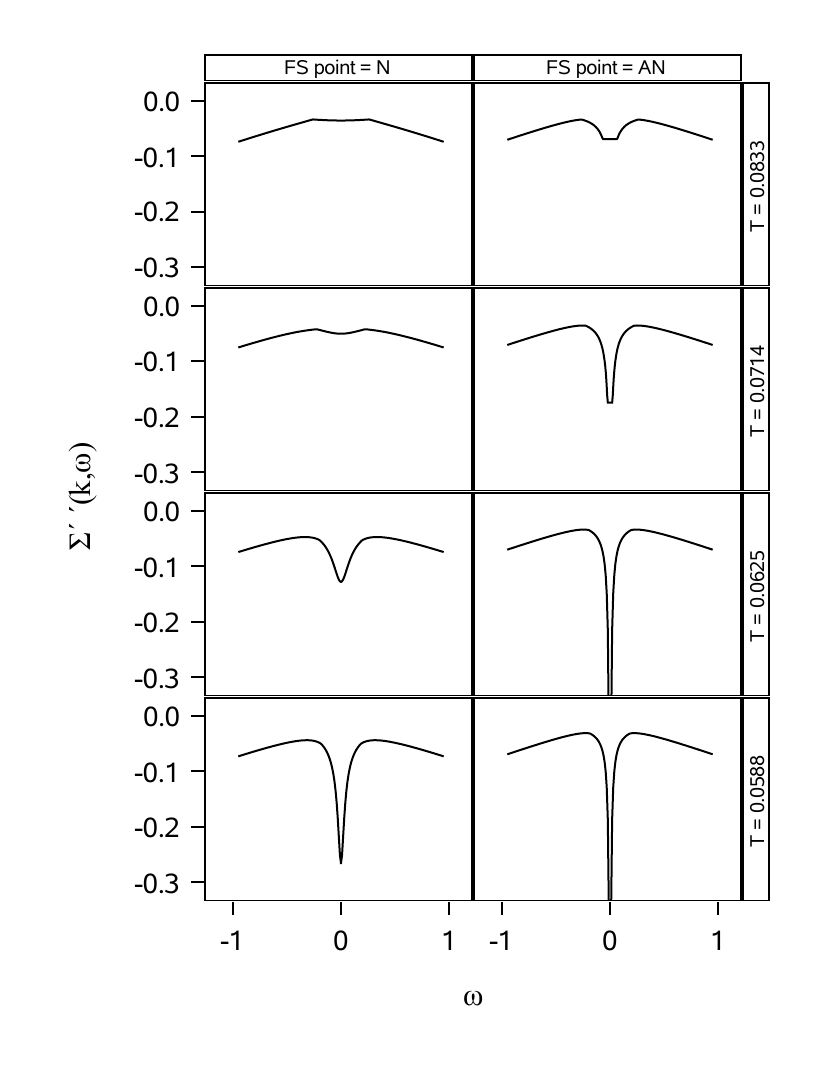} 
    \caption{Evolution of the imaginary part of the self-energy $\Sigma''(\mathbf{k}, \omega) $ with the temperature for nodal (left) and antinodal (right) points. The sharp minimum of this quantity at $\omega=0$ is the signature of the pseudogap.}
    \label{fig:real_freq_theory_im_Sigma}
\end{figure}

\begin{figure}
    \centering
    \includegraphics[width=\columnwidth]{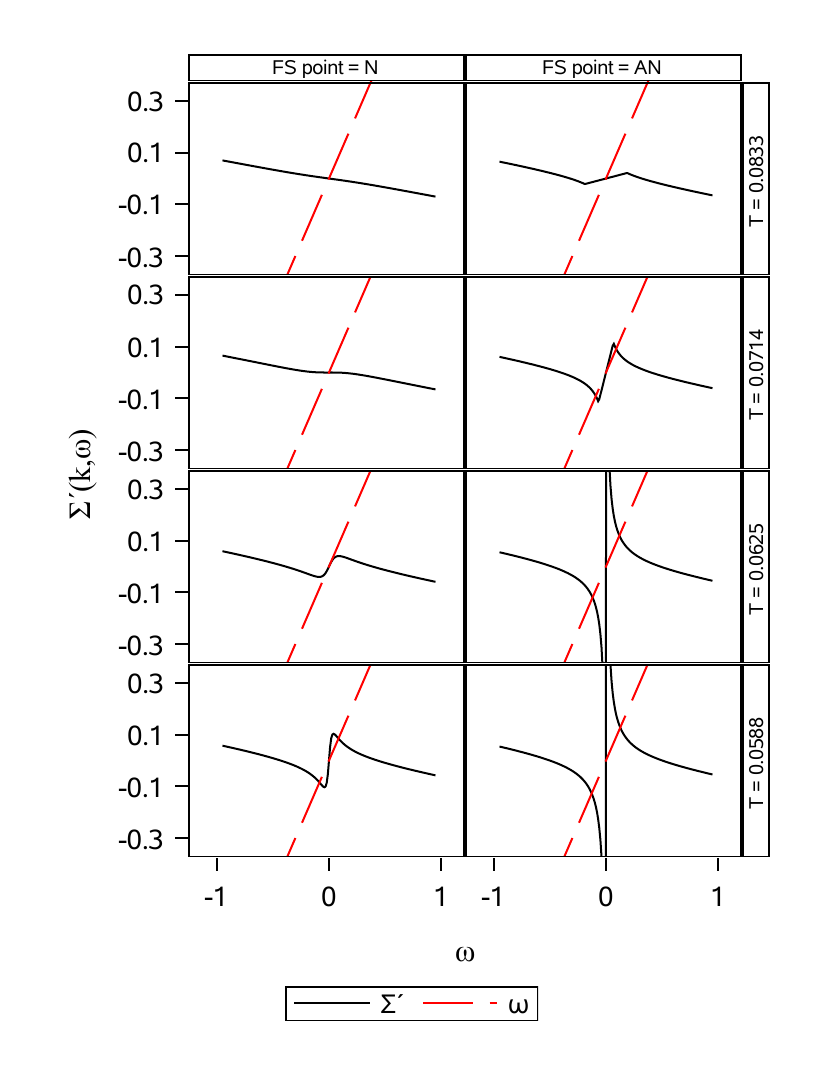} 
    \caption{Evolution of the real part of the self-energy $\Sigma'(\mathbf{k}, \omega) $ with the temperature for nodal (left) and antinodal (right) points. Intersections with line $\Sigma' = \omega$ determine the positions of the maximum in the spectral function $A(\mathbf{k}, \omega) $. The figure for AN point at $T=0.0833$ shows positive slope which is less than one. It is an example of the false quasiparticle (see text for details).}
    \label{fig:real_freq_theory_real_Sigma}
\end{figure}

	Panels in \fref{fig:real_freq_theory_A_k_om}, \fref{fig:real_freq_theory_im_Sigma}, \fref{fig:real_freq_theory_real_Sigma} show results for the spectral function $A(\mathbf{k}, \omega)=-2 \mathrm{Im} G((\mathbf{k}, \omega)$, the imaginary part of the self-energy $\Sigma''(\mathbf{k}, \omega) $, and the real part of the self-energy $\Sigma'(\mathbf{k}, \omega) $. The results are presented for four different temperatures below $T_{rc}=0.1$, and for the nodal and antinodal points on the FS. On the plots for the the real part of the self-energy $\Sigma'(\mathbf{k}, \omega) $, I also showed the line  $\Sigma'=\omega $. The intersections of this line with $\Sigma'(\mathbf{k}, \omega) $ are the solutions of equation for the poles of the Green function $\omega =\Sigma'(\mathbf{k}, \omega) $. The poles of the Green function correspond to the positions of the maximum in the spectral function $A(\mathbf{k}, \omega) $. When the slope of $\Sigma'(\mathbf{k}, \omega)$  at $\omega=0$ exceeds one, there are two maximum in the spectral function $A(\mathbf{k}, \omega) $ separated by the pseudogap.

In Ref.~\citenum{Schaefer2021} the pseudogap crossover temperatures for antinodal $T_{AN} =0.065$ and nodal points $T_{N} =0.0625$ were determined based on the Matsubara self-energy. As usual, the determination of the crossover is somewhat uncertain. The real frequency results give a little bit different perspective on the pseudogap \fref{fig:real_freq_theory_A_k_om}, show that the pseudogap already present at the AN point for $T=0.074$. At this temperature $\xi/\xi_{th\_vh} \approx 5$. On another hand, at the temperature $T=0.0625$ the pseudogap at the N point just start to develop. This is, especially, clear from the plot for the real part of the self-energy which shows that $\Sigma'(\mathbf{k_{N}}, \omega)$ is almost parallel to the line $\Sigma'=\omega$ at small $\omega$. At the temperature $T=0.0588$ the pseudogap is clearly developed at both antinodal and nodal points, but more pronounced at the AN point. At the opposite end of the temperature range at  $T=0.0833$, there is only one maximum in the spectral function for both the nodal and the antinodal points. However, the single-particle properties at the antinodal point are clearly abnormal at this temperature: the slope of the real part of the self-energy is positive and the imaginary part of the self-energy has a minimum instead of maximum at $\omega=0$. Similar, but less pronounced picture emerges at the temperature $T=0.074$ for the nodal point: the imaginary part of the self-energy has shallow minimum and the slope of the real part of the self-energy is almost zero. Although there is no pseudogap in this picture, the electronic state is clearly incoherent in this regime. I suggest to call it the false quasiparticle state. For the antinodal point it starts slightly below $T_{rc} $.

\begin{figure}
    \centering
    \includegraphics[width=\columnwidth]{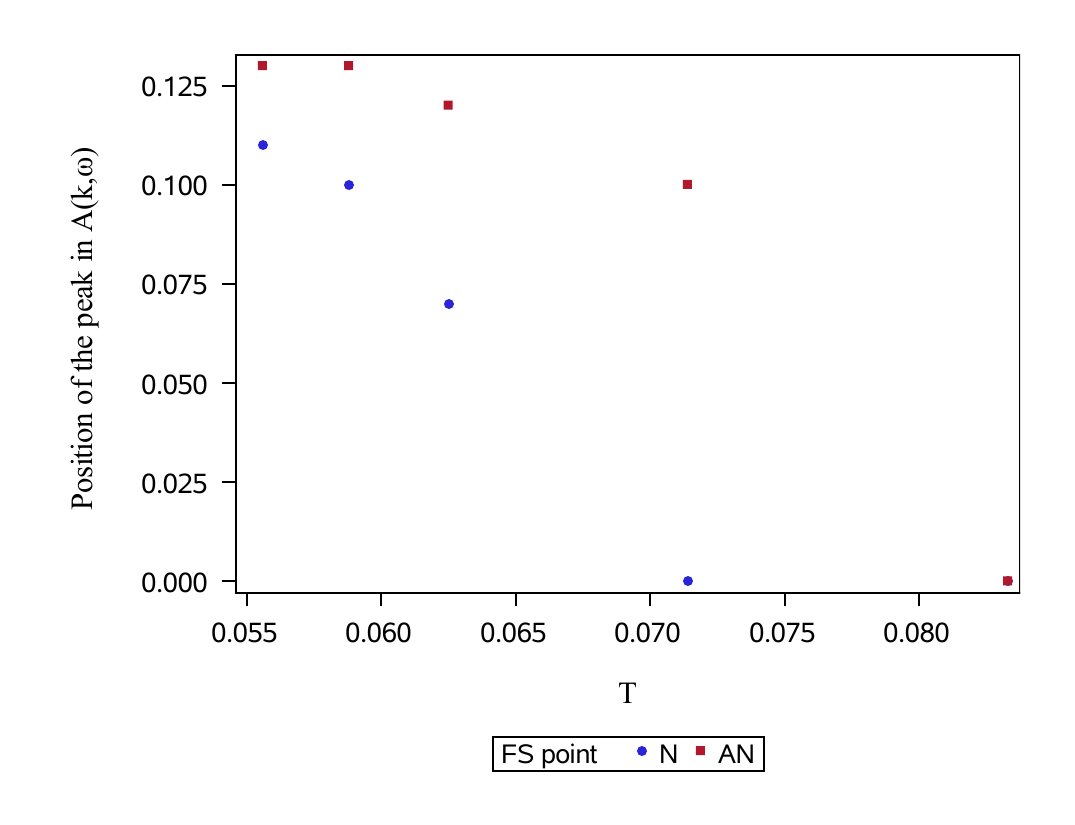} 
    \caption{Temperature dependence of the position of the maximum in the spectral function $A(\mathbf{k}, \omega) $ for nodal and antinodal points and $\omega \geq 0$. The position of the peak at the nodal point is always closer to the Fermi energy than at the antinodal point. This is expected based on the asymptotic results \eref{eq:peaks_position}, \eref{eq:peaks_position_R}.  }
    \label{fig:omega_of_peak_in_A_k_om}
\end{figure}

 The \fref{fig:omega_of_peak_in_A_k_om} shows positions of the peak in the spectral function at the N and AN points as a function of the temperature. The position of the peak at the nodal point is always closer to the Fermi energy than at the antinodal point. This is expected based on the asymptotic results for the position of the peaks in the spectral function \eref{eq:peaks_position}, \eref{eq:peaks_position_R}. The difference between positions of the peaks for antinodal and nodal points is temperature dependent and will disappear when $T \rightarrow 0$.

 We can see that the pseudogap behavior is significantly different at the antinodal point and at the regular point on the FS $v_F \sim 1$. As was pointed out in the previous section, the condition $w > v_F^2/(\pi T_{rc})$ defines an region close to the Van Hove point where the behavior of the single particle properties should be similar to the one at the Van Hove point. For the parameters considered here, the following section of the FS should behave similarly to the antinodal point: $\mathbf{k}=(\pi - k_y, k_y) $ with $ k_y < (1/2) (\pi T_{rc}/2)^{1/2} =0.2$.

 I conclude this section by pointing out, that the results in the section are fully applicable to the attractive Hubbard model at half filling and $U=-2$. As was mentioned earlier, the two models are equivalent via canonical transformation. In the attractive model, the pseudogap is due to critical thermal pairing. All results including criteria for the pseudogap and differences between behavior at the nodal and the antinodal points are the same.      

\section{Conclusion}
\label{sec:conclusion}

 In this paper I establish criteria and obtained analytical results for the pseudogap at the Van Hove  point on the Fermi surface in two dimensions. In two dimensions the mean field phase transition is suppressed due to thermal critical fluctuations (Mermin-Wagner theorem). It is replaced by a crossover to the renormalized classical regime (RC) with exponentially growing correlation length. It is in this regime that the pseudogap was predicted above the corresponding ordered state \cite{Vilk1996}, \cite{Vilk1997}.  I have also identified the state with false quasiparticles. In this state there is a single broad maximum in the spectral function but the self-energy is very abnormal.To describe pseudogap phenomena, I used phenomenological model which I validated by comparing with exact numerical Monte Carlo results \cite{Schaefer2021} for the Hubbard model at half-filling. The phenomenological model is applicable to both antiferromagnetic and pairing cases, and the condition for the pseudogap are the same. Specifically, the pseudogap appears when the correlation length is much larger than characteristic thermal wave length  $\xi >> \xi_{th}$. For the regular points on the FS, this is the de Broglie wavelength of an electron $\xi_{th}=\xi_{th\_db}=v_F/\pi T$ and for the Van Hove point it is $\xi_{th\_vh}=w/(\pi T)^{(1/2)}$. The latter is significantly shorter than the former and that is why the pseudogap appears at higher temperature and is more pronounced at the Van Hove point. For the Hubbard model at the intermediate coupling $ 2\leq U \leq 4$, the Van Hove characteristic length $\xi_{th\_vh}(T_{rc})$ is between one and two lattice spacing. 

 For arbitrary point on the FS, I suggest to use thermal characteristic length  $\xi_{th}= MAX(\xi_{th\_db}(T_{rc}), \xi_{th\_vh}(T_{rc})$. While the pseudogap condition is $\xi >> \xi_{th} $, in practice, the ratio  $\xi / \xi_{th} $ is not very large. In the experiments on electron doped materials \cite{motoyama2007spin} the antiferromagnetic pseudogap appeared when $\xi / \xi_{th} =2.6$. For the parameters of the Hubbard model in \secref{sec:Monte_Carlo} the pseudogap appears at $\xi / \xi_{th} \approx 4$. 

 The parts of the FS affected by the pseudogap are different for pairing and antiferromagnetic cases. In the former case, the peak in the susceptibility is at $\mathbf{q}=0$. Consequently, all points on the FS are affected by the pseudogap with the strongest effect at the Van Hove point. In the antiferromagnetic case the peak in the susceptibility is at $\mathbf{q}=\mathbf{Q}=(\pi, \pi)$. The singular behavior in the self-energy occurs on the shadow FS which is shifted by $\mathbf{Q}$ relative the real FS. The points where these two Fermi surfaces intersect ("hot" spots) are the points where the pseudogap develops in the antiferromagnetic case. In the electron doped High Tc materials the pseudogap occurs above antiferromagnetic state \cite{armitage2001anomalous}. It was shown in \cite{Kyung2004} that the pseudogap in these  materials can be explained quantitatively by antiferromagnetic thermal fluctuations in the RC regime. In the hole underdoped materials the pseudogap occurs above  superconducting state and smoothly transition to the true gap below Tc with the strongest effect at the Van Hove (antinodal) point \cite{Vishik2018}. The experiments also show that around the nodal point the gapless feature persists over wide range of temperatures. The latter observation is different from what is observed in the Hubbard model at the half-filling in the weak coupling limit. In that case, the pseudogap at the nodal point develops relatively soon after it develops at the antinodal point. Further studies are necessary to determine if the thermal classical  fluctuations in the RC regime are responsible for the pseudogap in the hole doped High Tc materials.
Another distinct possibility is the pseudogap due to strong coupling effects considered in Ref \cite{senechal2004hot}, \cite{sakai2023nonperturbative}.

\section*{Acknowlegments}
I thank A.-M.S. Tremblay for many stimulating discussions and suggestions. I am indebted to Thomas Sch\"afer for sharing data on the parameters for Ornstein-Zernike susceptibility.

\appendix

\section{Self-energy results for regular points on the FS}
\label{app:regular_k_F}

 In this appendix, I present results for regular points on the FS (($v_F \sim 1 $)). Most of the results shown below are from \cite{Vilk1996}, \cite{Vilk1997}. I reproduce them here for ease of reference.

 As I explained in \secref{sec:model}, for regular FS points the static approximation \eref{eq:sEnergy_st}  can be used for all frequencies in the RC regime. For the self-energy in the Matsubara frequencies representation, the expression is:

\begin{equation} \label{eq:sEnergy_st_Reg_M}
 \begin{split}
 \Sigma_{cl}(\mathbf{k}_{F},ik_n) & =\frac{g T A}{4 \pi  \sqrt{v_F^2 \xi^{-2} - k_n^2}} \\   
 &\ln \frac{ik_n+\sqrt{v_F^2 \xi^{-2} - k_n^2}}{ik_n-\sqrt{v_F^2 \xi^{-2} - k_n^2} }
\end{split}
\end{equation}

This expression of the complex variable reduces to two different functional expressions: one for $v_F \xi^{-1} > k_n$ and one for $v_F \xi^{-1} < k_n$. The latter region has been considered earlier in \cite{Vilk1996}, \cite{Vilk1997}. This is the region of temperatures where pseudogap occurs. The region $v_F \xi^{-1} > k_n$ will be for the first time presented here. It is exist in the RC regime when $ \gamma \xi^{-2} < \pi T < v_F \xi^{-1} $. I need this range of temperatures to compare with benchmark Monte Carlo results in \secref{sec:Monte_Carlo}.

Case $v_F \xi^{-1} > k_n$:

\begin{equation} \label{eq:sEnergy_st_Reg_M1}
 \begin{split}
 \Sigma_{cl}(\mathbf{k}_{F},ik_n) & =-i\frac{g T A}{2 \pi  \sqrt{v_F^2 \xi^{-2} - k_n^2}} \\   
 &\arctan \frac{\sqrt{v_F^2 \xi^{-2} - k_n^2}}{k_n} 
\end{split}
\end{equation}

Case $v_F \xi^{-1} < k_n$:

\begin{equation} \label{eq:sEnergy_st_Reg_M2}
 \begin{split}
 \Sigma_{cl}(\mathbf{k}_{F},ik_n) & =-i\frac{g T A}{4 \pi  \sqrt{k_n^2 -v_F^2 \xi^{-2}}} \\   
 &\ln \frac{k_n+\sqrt{k_n^2 -v_F^2 \xi^{-2}}}{k_n-\sqrt{k_n^2 -v_F^2 \xi^{-2}} }
\end{split}
\end{equation}

Let's now consider the case when correlation length $\xi$ is much larger than the de Broglie wavelength of an electron:

\begin{equation} \label{eq:Criteria_R}
\xi>> \xi_{th\_db}=\frac{v_F}{\pi T}
\end{equation}

This is the condition for the pseudogap at the regular point on the FS \cite{Vilk1996}, \cite{Vilk1997}. Expanding over small parameter $v_F \xi^{-1}/(\pi T)$ one obtains:

\begin{equation} \label{eq:sEnergy_st_M_R}
 \Sigma_{cl}(\mathbf{k}_{F},ik_n)=\frac{g T A}{2 \pi ik_n}\left[T\ln \xi +T \ln \left(\frac{2k_n}{v_F}\right)\right]
\end{equation}

Taking into account that in the RC regime correlation length grows exponentially $\xi=\tilde{\xi}_0 \exp(T_0/T)$ the \eref{eq:sEnergy_st_M_R} can be written as follows:

\begin{equation} \label{eq:sEnergy_st_R_M_Delta}
 \Sigma_{cl}(\mathbf{k}_{F},ik_n)=\frac{\Delta^2}{ik_n}\left[ 1+ \frac{T}{T_0} \ln \left( \frac{2k_n}{v_F \tilde{\xi}_0} \right)\right]
\end{equation}

Where gap parameter $\Delta$ is given by \eref{eq:Delta}.

I now present results for the self-energy in real frequency representation:

\begin{equation} \label{eq:sEnergy_st_Reg_R}
 \begin{split}
 \Sigma_{cl}(\mathbf{k}_{F},\omega) & = \frac{g T A}{4 \pi  \sqrt{\omega^2 +v_F^2 \xi^{-2}}} \\   
 &\left[ \ln \vline\frac{\omega+\sqrt{\omega^2 +v_F^2 \xi^{-2}}}{\omega-\sqrt{\omega^2 +v_F^2 \xi^{-2}} }\vline -i \pi\right]
\end{split}
\end{equation} 

At small $\omega < v_F \xi^{-1}$ the imaginary part of the self-energy $\Sigma_{cl}(\mathbf{k}_{F},0) \propto \xi $ and the slope of the  real part of the self-energy is positive and scales $ \xi^2 $. These are necessary conditions to have a pseudogap. 

At frequencies $\omega >> v_F \xi^{-1}$, the expression for the real part of the self-energy can be written using gap parameter as follows:

\begin{equation} \label{eq:sEnergy_st_R_R_Delta}
 \Sigma_{cl}(\mathbf{k}_{F},\omega)=\frac{\Delta^2}{\omega}\left[ 1+ \frac{T}{T_0} \ln \left( \frac{2 \omega}{v_F \tilde{\xi}_0} \right)\right]
\end{equation}
 
  Where gap parameter $\Delta$ is given by \eref{eq:Delta}. This form of the self-energy leads to two peaks in the spectral function, which are precursors of the quasi-particles in the ordered state. The positions of the peaks at low temperatures are given by:

\begin{equation} \label{eq:peaks_position_R}
 \omega_{peak}=\pm \Delta \sqrt{ 1+  \frac{T}{T_0} \ln \left( \frac{ 2 \Delta}{v_F \tilde{\xi}_0^{-1}} \right)}
\end{equation}

\section{Results for the regular contribution to the self-energy}
\label{app:regular_se}

In this appendix I present results for the regular contribution to the self-energy based on the  MFL model described in the \secref{sec:Monte_Carlo}.  I remind that the model for imaginary part of the self-energy is: 

\begin{equation}
    \Sigma''_r(\mathbf{k}, \omega)  =\begin{cases} 
      b|\omega| \,&  |\omega| \geq \omega_0=\pi T\\
      b \omega_0\,& |\omega| < \omega_0=\pi T\\
   \end{cases}
\label{eqn:im_part_r}
\end{equation}

  I start with the Matsubara self energy which I derive using Kramers-Kronig relation:
 	
\begin{equation} \label{eq:Kramers_Kronig}
 \Sigma(\mathbf{k},i k_n)=\frac{1}{\pi}\int \frac{\Sigma''(\mathbf{k},\omega)}{\omega - i k_n }
\end{equation}

The final result is:

\begin{equation} \label{eq:Sigma_M_r}
 \Sigma(\mathbf{k},i k_n)=i \frac{2b}{\pi}\left( \omega_0 \arctan\frac{\omega_0}{k_n} + k_n \ln \frac{\sqrt{\omega_{max}^2 + k_n^2}}{\sqrt{\omega_0^2 + k_n^2}}    \right)
\end{equation}

I now present the result for the real part of the self-energy:

\begin{equation} \label{eq:Sigma_R_r}
 \begin{split}
 \Sigma'(\mathbf{k}, \omega) & = \frac{b}{\pi}\left[ \omega \ln (\omega_{max}^2 -\omega^2 )\right] + \\ 
&  \frac{b}{\pi} \left[ (\omega_0-\omega) \ln(\omega_0-\omega)-(\omega_0+\omega) \ln(\omega_0 +\omega)  \right]
 \end{split}
\end{equation}

\bibliography{PG_VanHove.bib}

\end{document}